# Effects of electrical and optogenetic deep brain stimulation on synchronized oscillatory activity in Parkinsonian basal ganglia

Shivakeshavan Ratnadurai-Giridharan, Chung C. Cheung, and Leonid L. Rubchinsky

*Abstract*— Conventional deep brain stimulation (DBS) of basal ganglia uses high-frequency regular electrical pulses to treat Parkinsonian motor symptoms and has a series of limitations. Relatively new and not yet clinically tested optogenetic stimulation is an effective experimental stimulation technique to affect pathological network dynamics. We compared the effects of electrical and optogenetic stimulation of the basal ganglia on the pathological parkinsonian rhythmic neural activity. We studied the network response to electrical stimulation and excitatory and inhibitory optogenetic stimulations. Different stimulations exhibit different interactions with pathological activity in the network. We studied these interactions for different network and stimulation parameter values. Optogenetic stimulation was found to be more efficient than electrical stimulation in suppressing pathological rhythmicity. Our findings indicate that optogenetic control of neural synchrony may be more efficacious than electrical control because of the different ways of how stimulations interact with network dynamics.

## I. Introduction

DEEP brain stimulation (DBS) is a stimulation of the deep brain structures via implanted electrodes. It is used as a therapeutic procedure to treat symptoms of several neurological and neuropsychiatric disorders [1]. In particular, it is used to treat motor symptoms of Parkinson's disease (PD) by delivering high-frequency regular stimulation to targets, such as subthalamic nucleus (STN) [2].

The hypokinetic motor symptoms of PD are associated with excessive beta-band oscillations and synchrony in the basal ganglia and other brain parts [3,4]. While definite causative association between "pathological beta" and symptoms is a subject of debate (see [4]), many studies suggest that DBS in PD alleviates symptoms by suppressing pathological beta-band oscillatory synchronized activity [5,6]. However, the classical basal ganglia DBS in PD is not necessarily a very efficient procedure: it does not completely restore motor function and it has substantial side effects [7,8]. These effects may arise due to a high level of current delivered to motor circuits, so that it interferes with other motor functions and spreads to nearby cognitive pathways, which are not completely segregated [9].

An alternative way to stimulate neural circuits is via a relatively new technology of optogenetics. Optogenetic stimulation combines genetic and optical tools to stimulate specific neurons [10,11]. This specificity may be an important advantage of optogenetic stimulation with respect to electrical stimulation. However, optogenetic and electrical stimulations are also different in how they affect the electrical activity of neurons and networks. In particular, stimulation current delivered to neurons will depend on the state of a neuron in different ways. In optogenetics, stimulation-generated photocurrent depends on the transmembrane voltage. Thus, the stimulation becomes state-dependent.

The goal of this study is to explore how optogenetic stimulation compares with electrical stimulation in their network effects on elevated synchronized oscillatory activity. In this respect, suppression of excessive beta-band activity in the parkinsonian basal ganglia appears to us as an appropriate phenomenon to compare stimulations' network effects. We use a computational model of subthalamo-pallidal network of the basal ganglia to explore these phenomena.

Optogenetics is a relatively new experimental technique, not yet clinically tested. Our study will not have immediate clinical implications, but compares the network effects of two ways to control neural circuits. However, optogenetic stimulation of the basal ganglia has been implemented in rodents [12] and non-human primates [13]. If the issues of human implementation will be eventually resolved, there may be potential in clinical use of optogenetic DBS in PD. Moreover, optogenetic stimulation is a powerful experimental technique and this study highlights its potential efficacy and differences from electrical stimulation for controlling synchronized neural oscillations. We found that optogenetic and electrical stimulations interact with the network dynamics in different ways and that optogenetic stimulation may be more efficient in controlling pathological oscillatory synchronized dynamics of neural activity.

## II. Methods

### A. Network Model

There may be several different (not necessarily mutually exclusive) mechanisms of the parkinsonian beta-band activity. We adopt a conductance-based modeling of [14], which in

This work was supported by ICTSI/Indiana University Health–Indiana University School of Medicine Strategic Research Initiative.

S Ratnadurai-Giridharan was with IUPUI, Indianapolis, IN, USA. He is now with Burke Medical Research Institute, White Plains, NY, USA.

C.C. Cheung is with IUPUI, Indianapolis, IN, USA.

L.L. Rubchinsky is with Indiana University Purdue University Indianapolis and Indiana University School of Medicine, Indianapolis, IN, USA (e-mail: lrubchin@iupui.edu).

turn was based on [15] and considers STN and external Globus Pallidus (GPe). While this model is limited in several ways (for example, it excludes several brain areas, relevant for parkinsonian pathophysiology), it is based on the known anatomical and physiological data and incorporates rhythmicity mechanisms resulting from the recurrent excitation and inhibition in STN-GPe circuits [16]. Furthermore, this model reproduces patterns of synchronized oscillatory activity observed in Parkinsonian patients [17] and believed to be associated with hypokinetic motor symptoms.

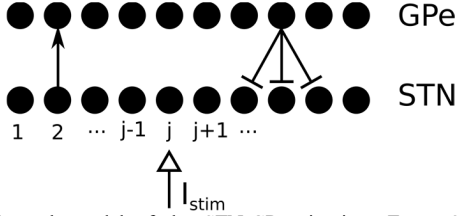

Fig. 1. Network model of the STN-GPe circuitry. Every STN neuron connects to a corresponding GPe neuron. Every GPe neuron connects to three neighboring STN neurons as shown in the figure.

From the dynamical systems standpoint, this model's phase space is similar to the one reconstructed from experiments not only at the vicinity of synchronization manifold, but at the periphery as well [14,18].

The model network has two arrays of neurons: ten GPe neurons and ten STN neurons (see Fig. 1). The network has circular boundary conditions. Each neuron is described by a conductance-based system of ordinary differential equations. The membrane potential obeys

$$C \frac{dV}{dt} = -I_L - I_K - I_{Na} - I_T - I_{Ca} - I_{AHP} - I_{syn} + I_{app}.$$

All the equations for membrane and synaptic currents and network parameters are available in [14]. Stimulation (described below) is applied to STN. Within this modelling framework, the expression of opsins would be over the dendrites and cell bodies.

*B. Electrical Stimulation*

Biphasic pulses of rectangular profile are used as stimulation currents. The waveform is structured as a 1 ms positive pulse followed by a 1ms negative pulse of equal amplitude, followed by *(p-2)* ms silent interval, where p is the period of the stimulation. This waveform may be described as

$$I_{elec} = A_{elec}(-1 + 2\Theta(p - 1 - mod(t,p))) \Theta(mod(t,p) - (p - 2)),$$

where $A_{elec}$ is the amplitude, p is the period, $\Theta(t)$ is the Heaviside step function.

*C. Optogenetic stimulation*

We consider Channelrhodopsin (ChR2) as a sodium light activated-channel to excite neurons by depolarization [19] and Halorhodopsin (NpHR) as a chlorine light-activated channel to suppress neuronal activity via hyperpolarization [20]. Below we describe the modeling of the light activation.

<u>Excitation.</u> We use a four-state transition rate model for the photocurrent kinetics following [21]. In particular, we use a model of ChETA, a ChR2 mutant with faster photocurrent dynamics. The photocycle transitions of ChETA are represented via two sets of intra-transition states: closed states $C_1$ and $C_2$ and open states $O_1$ and $O_2$ (Fig. 2a). Upon stimulation, there is a transition from the closed $C_1$ state to the open $O_1$ state, resulting in the photocurrent $I_{ChETA}$. From $O_1$, the channels may either relapse back to $C_1$ or transition to a secondary open state $O_2$ which has a lower photocurrent. When the light is turned off, channels transition from $O_1$ and $O_2$ to $C_1$ and $C_2$ respectively. There is a recovery period associated with $C_2$ to $C_1$ transition. Hence, successive stimulation pulses within short time lead to a lower photocurrent.

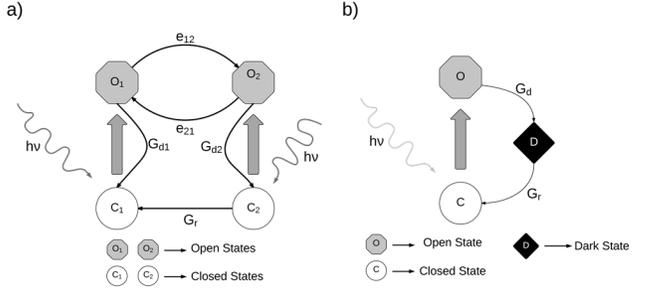

Fig. 2. Photocycles of Optogenetic stimulation. **a)** Four-state rate transition model to capture ChETA excitation dynamics. There are two open states (O1 and O2) and two closed states (C1 and C2). **b)** Three-state model for NpHR inhibition dynamics. There are open, closed, and transition dark states.

Let $c_1$, $c_2$, $o_1$ and $o_2$ denote the fraction of ChETA molecules in the $C_1$, $C_2$, $O_1$, and $O_2$ states respectively. The dynamics of the transitions between states was described as:

$$\dot{o}_1 = \varepsilon_1 uF(1 - c_2 - o_1 - o_2) - (G_{d1} + e_{12})o_1 + e_{21}o_2$$
$$\dot{o}_2 = \varepsilon_2 uFC_2 + e_{12}o_1 - (G_{d2} + e_{21})o_2$$
$$\dot{c}_2 = G_{d2}o_2 - (P_2 u + G_r)c_2$$
$$\dot{u} = (S_0(\phi) - u)/\tau_{ChR2}$$

Variable $c_1$ is excluded because $c_1 + c_2 + o_1 + o_2 = 1$. The parameters $\varepsilon_1$, $\varepsilon_2$, $G_{d1}$, $G_{d2}$, $e_{12}$, $e_{21}$, $G_r$ represent the transition rates. $\tau_{ChR2}$ is the activation time of the ChETA ion channel (1.5855 ms). The function $u$ captures the temporal kinetics of the conformational change in protein [21]. The number of photons absorbed by ChETA molecule per unit time is given by $F = \sigma_{ret}\varphi/w_{loss}$, where $\sigma_{ret}$ is the retinal cross-section (1.2 x $10^{-20}$ m$^2$) [22] and $w_{loss}$ is the loss of photons due to scattering and absorption. $\varphi$ is the photon flux per unit area: $\varphi = \lambda A/hc$, where $\lambda \approx 480$ nm is the wavelength of stimulating blue light, $A$ is the intensity of light stimulation, $h$ is Planck's constant, and $c$ is the speed of light. The function $S_o(\phi) = 0.5(1+\tanh(120(\phi-0.1)))$ is a sigmoidal function and $\phi(t) = \Theta(mod(t,p) - t_{off})$ describes the stimulation protocol, where $p$ is the period of stimulation, and $t_{off}$ is the time per cycle when the stimulation is turned off (so that $t_{on} = p - t_{off}$ is a pulse duration). $\Theta$ is the Heaviside step function. We consider light pulses as $A(t) = A_{light}\phi(t)$ where $A_{light}$ is the constant characterizing the light intensity.

The ChETA photocurrent is given by
$$I_{ChR2} = g_{ChR2}(V - V_{Na})(O_1 + \gamma O_2),$$

where $g_{ChETA}$ is the maximal conductance of the ChETA channel in the $O_1$ state, $V_{Na}$ is the reversal potential, and $\gamma$ represents the ratio of conductances in $O_1$ and $O_2$ states. See [21] for parameter values.

<u>Inhibition.</u> We develop a three state model to reproduce the photocurrent dynamics observed experimentally [11,23]. We use a three state model, a minimal requirement in order to capture the photocurrent dynamics of Halorhodopsin channels.

Additionally, NpHR exhibits at least three states during light activation/inactivation [24]. Our model assumes three different states for the NpHR channels: closed state C, open state O, and the desensitized state D. Let c, o, and d, denote the fraction of NpHR channels in these states. We describe the dynamics of the transitions between these states as:

$$\dot{o} = \varepsilon F \phi(t)(1 - o - d) - G_d o$$
$$\dot{d} = G_d o - G_r d$$

where $c + o + d = 1$, so that $C$ is eliminated from the equations. As above, $F = \sigma_{ret}\varphi/w_{loss}$, $\varphi$ is the photon flux per unit area: $\varphi = \lambda A/hc$, where $\lambda \approx 570$ nm is the wavelength of stimulating yellow light. Light pulses are applied as above: $A(t) = A_{light}\phi(t)$. The NpHR photocurrent is given by

$$I_{NpHR} = g_{NpHR}(V - V_{Cl})O,$$

where $V_{Cl}$ is the reversal potential of the chlorine channel.

We estimated the parameters $G_d$, $G_r$, $\varepsilon$, and $g_{NpHR}$ by using experimental results [11,23] to construct an empirical curve that represents NpHR photocurrent dynamics. The empirical curve that fits the temporal profile of experimentally observed photocurrents can be estimated similar to [21] using:

$$I_{NpHR}^{emp} = I_{peak} \left\{ \begin{array}{l} \left(1 - e^{-\frac{t-t_{on}}{\tau_{rise}}}\right)\theta(t - t_{on})\theta\left((t_{on} + t_p) - t\right) \\ + \\ \left(R + (1-R)e^{-\frac{(t-(t_{on}+t_p))}{\tau_{in}}}\right)\theta\left(t - (t_{on} + t_p)\right)\theta(t_{off} - t) \\ + \\ R\, e^{-\frac{(t-t_{off})}{\tau_{off}}}\theta(t - t_{off}) \end{array} \right\}$$

where $I_{peak} = 42.8$ pA is the peak photocurrent, $R = 0.8505$ is the ratio of steady to peak current values, $t_{on} = 50$ ms, $t_{off} = 150$ ms, $t_p = 1$ms is the time taken to reach the peak photocurrent value, $\tau_{rise} = 0.5$ ms is the rise time constant, $\tau_{in} = 2$ms is the time constant for the decay from the peak photocurrent to steady state, $\tau_{off} = 6.9$ ms is the decay constant from steady state photocurrent to zero, $\Theta$ is the Heaviside step function.

To estimate $G_d$, $G_r$, $\varepsilon$, and $g_{NpHR}$, we consider a cost function $L = \|I_{NpHR} - I_{NpHR}^{emp}\|$, where $\|\cdot\|$ is $L^2$ norm. We minimized $L$ using the Nelder-Mead simplex method [25] and obtained values $\{G_d, G_r, \varepsilon,$ and $g_{NpHR}\} = \{0.6518$ ms$^{-1}$, $0.1385$ ms$^{-1}$, $0.4296$ ms$^{-1}$, $2.4002$ mS/cm$^2\}$.

### D. Characterization of Network's Dynamics

The network model used here has been studied in [14] without stimulation within the two-dimensional parameter space of $(g_{syn}, I_{app})$, where $g_{syn}$ is the strength of the synaptic connections from GPe to STN and $I_{app}$ is the constant current applied to GPe to model average impact of striatopallidal inhibition. These parameters are thought to be modulated by dopamine, which degenerates in PD. Larger values of $g_{syn}$ and smaller values of $I_{app}$ correspond to a low-dopamine state [14].

To quantify the beta activity in the network, we use spikes as it was done with experimental [17] and model STN [14] data. Parkinsonian state in the model is anti-phase synchronous [14,15]. Mean field will be very small for the strong anti-phase synchrony. So to capture the oscillations and synchrony in the beta band in the model, we observe the mean field for each of the two anti-phase clusters separately from each other. This is not an arbitrary choice, it is rather induced by the nature of the observed synchronized dynamics. This results in the following measure of the beta-band activity:

$$\beta_{act} = \frac{1}{n}\left[Var\left(\sum_{i=1}^{n/2} s_{2i-1}\right) + Var\left(\sum_{i=1}^{n/2} s_{2i}\right)\right]k,$$

where $s_i$ is the filtered spiking signal of the $i^{th}$ STN neuron,

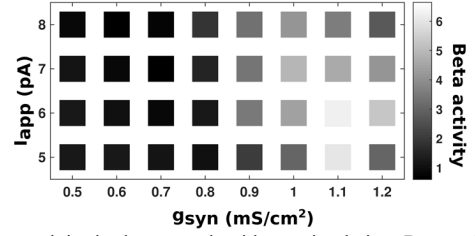

Fig. 3. Beta activity in the network without stimulation. By varying $I_{app}$ and $g_{syn}$ we achieve different levels of beta activity as seen in the figure. In general, lower levels of beta activity can be achieved with lower $g_{syn}$ values and higher $I_{app}$ values.

$Var$ is the variance function, and k is a scale factor, chosen here to be $10^5$ to get the values of $\beta_{act}$ around 1 to roughly correspond to where the lower end of the values from experimental data would start [14]. $\beta_{act} \sim 0$ if the beta-band oscillations are very weak or if the individual neurons are not synchronized in the beta band.

The quantification of the beta activity in the unstimulated network is presented in Fig. 3. Right lower (upper left) corner has more (less) beta activity and is a synchronized (less synchronized) state. Park et al. [14] showed that the dynamics of the model in this parameter domain exhibits synchronous patterns similar to those observed in the recordings from the parkinsonian patients [17]. To differentiate between the healthy and parkinsonian state we assume the threshold value of $\beta_{act}$ is 1. Network with $\beta_{act}$ above this threshold is considered to be in a Parkinsonian state. While there is no sharp transition between a healthy and parkinsonian state, we need to select the threshold to see when beta activity is suppressed to the desired level. The results persist qualitatively, when we varied the threshold by 25%.

### E. Quantification of stimulation to suppress beta-band activity

In electrical stimulation we varied the stimulation amplitude $A_{elec}$ from 5 to 100 pA with increment of 5 pA and from 100 to 150 pA with increments of 10 pA. In optogenetic stimulation, the light intensity was varied with $A_{light} = \{2,5,7,10,17,25,37,50\}$ mw/mm$^2$ and the off-time, $t_{off}$, for light per stimulation period was varied from $(p-1)$ ms to $(p/2)$ ms in steps of 1 ms, where $p = 10$ ms. All stimulations are done at the frequency of 100Hz.

To compare electrical and optical stimulation, we use RMS current delivered to the network averaged over time T:

$$I_x^{RMS} = \frac{1}{10}\sum_{i=1}^{10}\left(\frac{1}{T}\int_0^T I_{x,i}^2 dt\right)^{1/2},$$

where $I_x$ can be $I_{elec}$, $I_{ChETA}$ or $I_{NpHR}$.

To compare the efficacy of optogenetic stimulations with that of electrical stimulation, we consider the minimum $I^{RMS}$ needed to suppress the beta activity below the threshold. The minima are computed over varied stimulation intensities and light pulse durations. These quantities are normalized:

$$\eta_{ChR2} = \frac{I_e^{RMS} - I_{ChR2}^{RMS}}{I_e^{RMS}},$$

$$\eta_{NpHR} = (I_e^{RMS} - I_{NpHR}^{RMS})/I_e^{RMS}$$

If $\eta > 0$, the RMS current used by optogenetic stimulation to suppress beta is less than that of electrical one. We consider this as optogenetic stimulation being more effective than electrical: it reaches the same suppression effect with smaller currents to (and thus, perhaps, smaller impact on) the neurons.

III. RESULTS

A. Impact of different stimulation types on neuronal and network dynamics

Without stimulation, activity of the STN ranges from strong anti-phase synchronized bursting to less synchronized activity with less prominent bursting, depending on the values of network parameters $g_{syn}$ and $I_{app}$. The activity of one STN neuron from the network in the case of partially synchronized network dynamics ($I_{app}$ = 7 pA, $g_{syn}$ = 0.9 mS/cm$^2$) is presented at Fig. 4a. The interspike interval (ISI) histogram in Fig. 4a characterizes this spiking/bursting activity. The corresponding network activity is shown in Fig. 5a. There is a tendency of anti-phase burst synchrony with irregular and mixed spiking/bursting activity. In this case, $\beta_{act}$ = 3.4 (see Fig. 4), which may be within the range of PD neural activity [14].

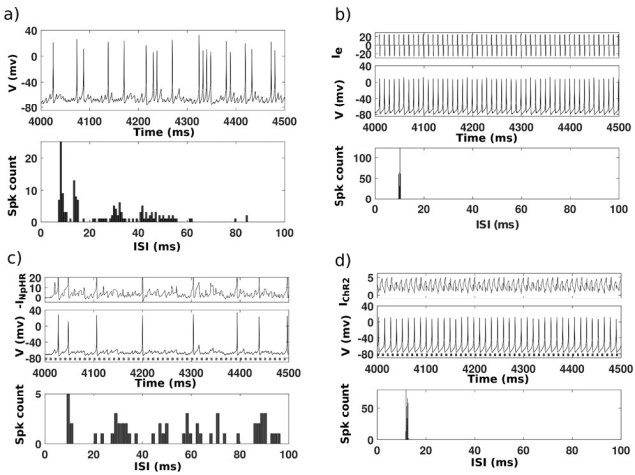

Fig. 4. Activity of an STN neuron for $I_{app}$=7, $g_{syn}$=0.9. **a)** Unstimulated STN neuron shows spiking and bursting. The ISI diagram indicates both high frequency activity and beta-frequency activity. **b)** STN activity in electrical stimulation aligns with stimulation: stimulation current, almost periodic spiking activity of STN, and ISI distribution. **c)** STN activity in optogenetic inhibition: inhibitory stimulation current, resulting suppressed neuronal activity, and varied timescales present at ISI distribution. Neurons are mostly hyperpolarized and the overall activity is very low. **d)** STN activity during optogenetic excitation: stimulation current, periodic spiking activity, and ISI distribution. The neurons are driven to fire in a regular high-frequency pattern with a period slightly longer than stimulation period. Light stimulation pulses are shown in c) and d) as rectangles below voltage trace.

We now present the neuronal and network dynamics for three stimulation types using typical stimulation parameters. In all cases the stimulation leads to suppression of the beta activity. Systematic study of the network dynamics in dependence on stimulation parameters is presented below.

A strong 100 Hz electrical stimulation current drives STN neurons to produce regular tonic firing at ≈100 Hz (Fig. 4b). The ISI histogram shows no noticeable beta-band activity. The STN network dynamics (Fig. 5b) shows that STN neurons are synchronized at 100 Hz. Thus there is no beta activity.

Inhibitory optogenetic stimulation presents a very different dynamic: it decreases the firing rate and suppresses burstiness (Fig. 4c). ISI histogram shows a range of spiking rates, but the spike counts are extremely low due to inhibitory effect of the stimulation. The network activity (Fig. 5c) is sparser without stimulation. There is no marked beta-band activity.

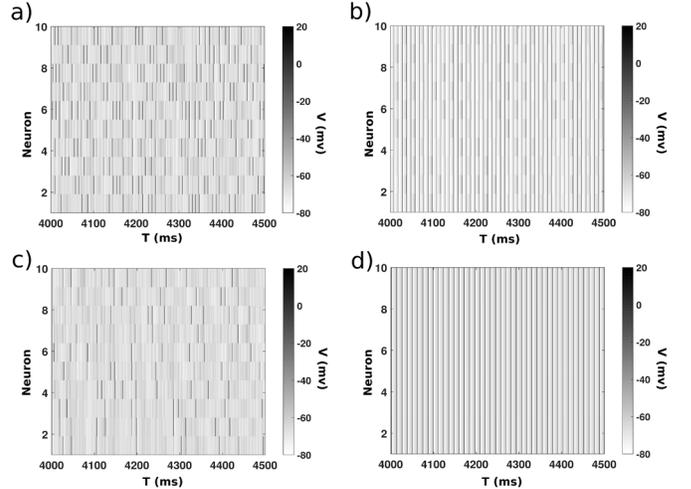

Fig. 5. STN voltage raster plot depicting network activity. Dark lines correspond to spikes. **a)** Network activity without stimulation shows a mix of spiking and bursting with partial anti-phase synchrony. **b)** Electrical stimulation causes the network to synchronously fire at the frequency of stimulation. **c)** Network activity is sparse and unsynchronized under optogenetic inhibition. **d)** Optogenetic excitation causes the network to spike synchronously at the frequency lower than stimulation frequency.

Optogenetic excitation at 100Hz has similar in effect to electrical stimulation. The STN neurons fire tonically (Fig. 4d). The network activity is synchronous high-frequency spiking outside of the beta-range, resulting in $\beta_{act}$ near zero (Fig. 5d). However, unlike electrical stimulation, the neurons are firing at ≈ 85 Hz. When successive light pulses occur within short time spans of 10 ms, the optogenetic channels may not have completely recovered and are not in their most conductive state. Hence unlike in electrical stimulation, the stimulation current appears like ongoing fluctuations around some positive value rather than brief regular pulses. As a result, the spiking is not 1:1 locked to stimulation.

Sufficiently strong stimulation (as in the three examples above) abolishes beta-band activity in all three stimulation types, although in different ways. Electrical stimulation and optogenetic excitation are similar and suppress beta activity by shifting neuronal firing patterns to synchronous high frequency spiking. Optogenetic inhibition suppresses activity overall. Intermediate stimulation strength leads to a smaller suppression of the beta-band activity in individual neurons. However, it may also exert some desynchronizing effect, which also contributes to the lowering of $\beta_{act}$.

B. Varying parameters of stimulation

Here we examine how the intensity and duration of stimulation pulses affect beta activity measured by $\beta_{act}$. We vary intensity and pulse duration and observe how $\beta_{act}$ changes in networks with three different values of ($g_{syn}$, $I_{app}$): (7,0.9) is moderate beta activity ($\beta_{act}$ = 3.39) without stimulation as in the examples above; (6,1) is higher levels of beta activity without stimulation ($\beta_{act}$ = 4.46); (8,0.8) is lower levels of beta activity ($\beta_{act}$ = 1.86) without stimulation.

Increasing amplitude of electrical stimulation does not necessarily cause an immediate decrease in beta activity (Fig. 6a). Weak stimulation may slightly elevate beta activity. However, further increase of the amplitude leads to a steep decrease in beta-activity. This increase for weak stimulation is attributed to the fact that with weak electrical stimulation, the neurons are not driven to fire at the stimulation frequency. Instead, weak electrical stimulation may enhance existing network dynamics, thus potentially reinforcing beta activity. If stimulation amplitudes increase beyond a certain range the beta activity decreases significantly. Fig. 6a suggest that this is true independently of $g_{syn}$ and $I_{app}$.

The increase of light intensity in optogenetic inhibition leads to a sharp decrease in the beta activity (Fig. 6b). This is consistent with the idea that stronger light leads to larger hyperpolarizing currents in the stimulated neuron, effectively suppressing neuronal and network activity. The dependence of $β_{act}$ on the light intensity is similar for all three cases of network parameters.

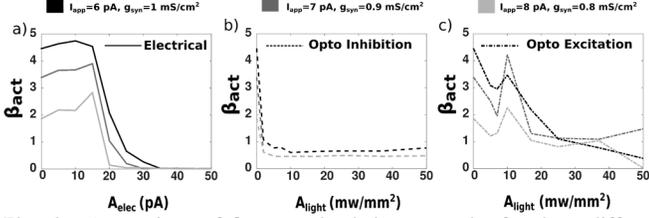

Fig. 6. Comparison of $β_{act}$ vs. stimulation strengths for three different networks (determined by $I_{app}$ and $g_{syn}$, illustrated in black, dark gray, and light gray colors). **a)** electrical stimulation, **b)** optogenetic inhibition, and **c)** optogenetic excitation.

For optogenetic excitation, the relationship between light intensity and beta activity is more complex. $β_{act}$ does not exhibit monotonous dependence on light intensity $A_{light}$ (Fig. 6c). While initial increase of light intensity decreases beta activity in the network, $β_{act}$ eventually exhibits a substantial peak. However, as $A_{light}$ further increases ($A_{light} > 15$ mw/mm$^2$), beta activity is low. The observed fluctuations in $β_{act}$ vs. $A_{light}$ may be due to weak photocurrents pushing network activity in a direction where beta activity is increased. And again, the overall dependence of beta activity on stimulation intensity does not vary much for different values

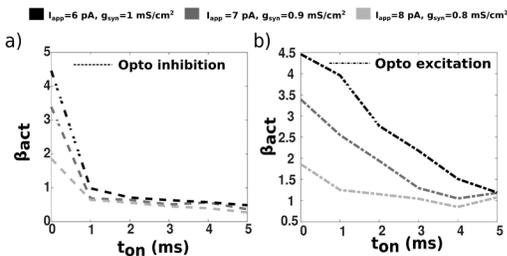

Fig. 7. Beta activity $β_{act}$ vs. light pulse durations. Different shades of grey correspond to different values of parameters $I_{app}$ and $g_{syn}$ as indicated. a) optogenetic inhibition (dashed lines), b)optogenetic excitation (dotted lines). Light intensities for both inhibition and excitation are fixed at 17 mw/mm$^2$.

of network parameters.

While clinically effective conventional electric DBS utilizes stimulus with very short pulse duration, short pulses may not necessarily be the most effective for optogenetic stimulation. Thus, for optogenetic stimulation we also study the response of the beta activity to different pulse durations $t_{on} = p − t_{off}$, where $p$ is period of stimulation. We again consider three different sets of network parameters. Inhibitory stimulation leads to a strong suppression of beta activity even with short pulse durations (Fig. 7a). Excitatory stimulation exhibits more gradual suppression (Fig. 7b). However, overall, increasing the durations of the light pulse leads to the decrease in beta activity.

### C. Comparing efficacy of Optogenetic and Electrical Stimulations

Stimulation parameters varied above, intensity of the stimulation and duration of stimulus, may be directly varied in experiment. However, what ultimately affects neural activity is the current a neuron experiences from stimulation. In optogenetic stimulation it also depends on the state of the stimulated neurons and interactions between stimulation and neuronal dynamics. Thus, we explore how $β_{act}$ depends on the stimulation current experienced by neurons (which is not necessarily directly proportional to the stimulation intensity). To do this we measure the RMS current delivered to a STN neuron on average for each of the stimulation types for different stimulation parameters.

We first explore how beta activity depends on the stimulation current for different stimulation types. We vary stimulation parameters creating a large array of stimulations with different parameter values, for which we compute resulting RMS stimulation currents. These currents are not directly controlled in experiment, but we have a reasonably large span of values to compare different stimulations types. This comparison is presented in the Fig. 8 for the same three

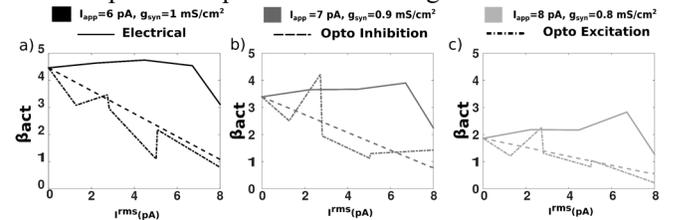

Fig. 8. Beta activity vs. RMS current induced by different stimulation types for three different STN-GPe networks. Subplots **a)**, **b)**, and **c)** correspond to networks with different values of $I_{app}$ and $g_{syn}$ (as indicated by different shades of grey). The types of stimulation are electrical (solid line), optogenetic inhibition (dashed line), and optogenetic excitation (dotted line).

sets of network parameter values.

Beta activity exhibit non-monotonous, but generally decaying dependence on RMS stimulation current (Fig. 8). Electrical stimulation RMS current is proportional to stimulation intensity, so that electrical stimulation curves in Fig. 8 follow Fig. 6a. Optogenetic excitatory stimulation exhibit non-monotonous dependence on RMS stimulation current (cf. non-monotonous dependence of $β_{act}$ on $A_{light}$, Fig. 6c). Optogenetic inhibitory stimulation exhibits gradual decrease of $β_{act}$. Although beta activity depends on the light intensity in inhibitory and excitatory stimulations in a very much different ways (Fig. 6b,c), this difference is probably due to different currents elicited by the stimulations. The dependence of $β_{act}$ on RMS current is not very different for inhibitory and excitatory optogenetic stimulations (Fig. 8).

If we can consider RMS current as a proxy for efficacy, we can compare the efficacies by comparing $β_{act}$ vs. $I^{rms}$ in the Fig. 8. Optogenetic inhibition provides consistent suppression, while optogenetic excitation is sometimes more and sometimes less effective that inhibitory one. However beta activity for electrical stimulation is mostly above that of the optical stimulations. Thus, electrical stimulation may be less effective than optogenetic stimulations.

To explore this issue systematically, we find the minimal (for different stimulation parameters) value of $I^{rms}$ required to bring $\beta_{act}$ below the threshold. We then compare relative efficacy of electrical and optogenetic stimulations by comparing the amount of stimulus (RMS current) between optogenetic and electrical input. The comparisons are done using $\eta_{NpHR}$ and $\eta_{ChETA}$, measuring how much more or less current is required by optogenetic excitation and inhibition relative to electrical stimulation to suppress beta activity. $\eta<0$ indicates that electrical stimulation is more efficacious (requires less current) than optogenetic stimulation, $\eta>0$ points to the efficacy of optogenetic stimulation.

The studies of synchrony patterns in parkinsonian basal ganglia [14,17], which developed the network models we use, cannot precisely specify the values of parameters $g_{syn}$ and $I_{app}$ for Parkinson's disease. So we explore a relatively large range of these parameters, which according to [14] is likely to include parkinsonian activity. We compute the minimal possible $I^{rms}$ to suppress beta activity for each stimulation type for a range of values of $g_{syn}$ and $I_{app}$ (Fig. 9).

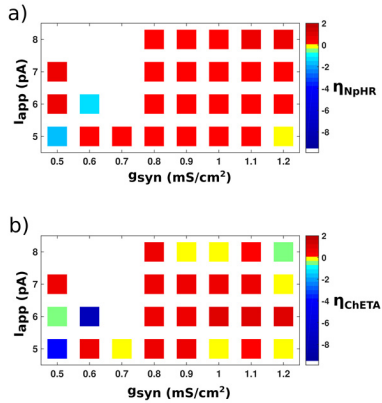

Fig. 9. Comparison of efficacies of electrical and optogenetic inhibitory(a) and excitatory (b) stimulation. $\eta > 0$ indicates efficacy of optical stimulation over electrical stimulation (less stimulation current is needed to reach the desired effect).

In Fig. 9a, for a wide range of network dynamics determined by $g_{syn}$ and $I_{app}$, $\eta_{NpHR} > 0$. Thus optogenetic inhibition suppresses beta activity with smaller current than electrical stimulation. Several points ($g_{syn}$, $I_{app}$) have no $\eta$ values assigned because unstimulated dynamics satisfies the condition $\beta_{act} < 1$. Overall, since $\eta_{NpHR}$ is mostly positive, optogenetic inhibition in the model network is more effective than electrical stimulation for most of the parameter values.

In optogenetic excitation (Fig. 9b) majority of the squares have $\eta_{ChETA} > 0$, meaning optogenetic excitation is more efficient than electrical stimulation for the majority of the parameter values considered here. The number of points with $\eta_{ChETA} > 0$ is smaller than the number of points with $\eta_{NpHR} > 0$, so optogenetic inhibitory stimulation may be viewed as to be more efficacious in general than optogenetic excitatory one. However, if we also consider $\eta \approx 0$ cases, then even optogenetic excitatory stimulation is no less (if not more) efficacious than electrical stimulation for most of the parameter values. Thus, for most of the parameter values, the stimulation currents in optogenetic stimulation (both excitatory and inhibitory) are smaller than currents in electrical stimulation.

The results presented above are for the stimulation frequency 100 Hz. Although we did not perform comprehensive studies of other frequencies, we observed qualitatively similar results for 147 Hz stimulations.

## IV. Discussion

### A. Electrical vs. optogenetic stimulation: efficacy and mechanisms

Using computational modeling, we observe that electrical DBS and two types of optogenetic DBS (excitatory and inhibitory) can decrease synchronized beta activity commonly associated with hypokinetic parkinsonian symptoms. Higher strength of stimulation generally leads to larger suppression of the beta-band synchronized oscillatory activity and stimulation types are similar to each other in this sense. However, the actions of different stimulation types on the beta activity may differ from each other.

Electrical stimulation and optogenetic excitation have somewhat similar network effects. Weak intensities of these stimulations may enhance beta activity. This non-monotonicity is also observed for the resulting effective stimulation current of optogenetic excitation (which is not necessarily proportional to light intensity due to the dynamics of photosensitivity). However, increasing stimulation intensities drive beta activity down. As stimulation strength is increased, they both synchronize the network at high (non-beta) frequencies in a tonic spiking dynamics, which effectively suppresses beta activity. Optogenetic inhibition presents a different mechanism of suppression. It reduces neural activity. Strong light suppresses all (not only beta) activity. This is not unlike lesion therapy in PD [26]. Interestingly, inhibitory optogenetic STN stimulation alleviated motor symptoms in a rodent model of PD [27] (although an earlier study failed to find a similar effect [28]).

We found that optogenetic inhibition usually requires less effective current than electrical stimulation to achieve beta suppression. Optogenetic excitation, while not as generally efficacious as inhibition, still usually requires less effective current than electrical DBS to suppress beta activity. Even in the cases where optogenetic excitation is not more efficacious than electrical DBS for beta activity suppression, the required amount of optical DBS effective current is usually close to the effective current needed by electrical DBS. Thus, our results suggest that optogenetic stimulation may introduce less or equal amounts of effective currents to a neuron and still achieve sufficient beta activity suppression.

Possible explanation is that optogenetic current depends not only on the light intensity, but also on the state of the neuron. Kinetics of photosensitive currents may serve as a feedback control for the stimulation current. Our results suggest this feedback may lead to efficacy of optogenetic stimulation being higher than efficacy of conventional electrical DBS. The importance of a high efficacy lies in the possibility that side effects of DBS may be reduced with lower stimulation current.

In relation to this, it is interesting to recall computational studies of feedback-based electric stimulation, which may require currents smaller than that of conventional open-loop DBS [29-31]. We considered optogenetic stimulation in an open loop protocol (which makes it easier to implement), however, the stimulation current is regulated by the state of the neurons, which is essentially a feedback mechanism. Unlike the closed-loop DBS, this feedback is not designed to suppress beta activity, but happens to be effective.

*B. Limitations of modeling and robustness of the stimulations' efficacy comparison*

Our modeling (as any other neural modeling) is limited in many ways and lacks a large number of anatomical and physiological features of the real brain. Small size and homogeneity of the network exclude spatial effects of stimulation and potential cellular and synaptic heterogeneity. Neurons have complex geometry, which may affect the stimulation outcome. The model electrical stimulation is essentially an intracellular stimulation. The stimulation in the model is homogeneous, while the actual distribution of the electrical field or light is spatially inhomogeneous. Electrical stimulation will differentially activate all neural elements including passing fibers, while optogenetic stimulation will affect the cell types expressing the opsins. So, our modeling results should be interpreted with appropriate caution. They are not as much definite, as suggestive that the dependence of the stimulation current on the state of the neuron in the optical stimulation may lead to substantial differences from and better efficacy than in electrical stimulation.

There is the question of whether the model correctly represents beta activity. We suppose the model captures some of the pathological rhythmicity mechanisms related to the recurrent excitation and inhibition in STN-GPe networks [16]. Moreover, the model reproduces experimentally observed patterns of beta activity [14]. The matching of the experimental and model synchrony patterns ensures similarity of the actual and the model phase spaces ([18,32,33]). So, the considered model may be dynamically adequate for the study of some mechanisms of beta suppression.

There is also some uncertainty regarding the causal connection between the beta activity and parkinsonian hypokinetic symptoms. However, the presently used therapies, including electrical DBS do exert beta-suppressive effects (see discussion of different experiments in [4]).

The value of the beta activity suppression threshold was chosen based on the recordings from the Parkinsonian patients – a necessarily imprecise approach (we do not know basal ganglia activity in the healthy humans). The suppression of the synchronized beta activity is not monotonous with the stimulation strength. We considered several threshold values, below which beta activity needs to be suppressed. For larger thresholds, the results are getting ambiguous (largely because little if any stimulation is needed to bring the activity below the threshold). However, as the threshold is decreased, we observed outcomes similar to what we described above.

## V. CONCLUSIONS

Our results suggest that optogenetic stimulation may be an effective alternative to conventional electrical DBS in suppression of pathological synchronized oscillations in parkinsonian basal ganglia. Optogenetic inhibition appears to be almost always more effective than electrical DBS. Optogenetic excitation, while not as efficacious as inhibition, may still be a competitive alternative to electrical DBS.

Our computational results appear to be robust enough to point to the potential of optogenetic DBS. The optogenetic stimulation may be remote from clinical practice for various reasons. However, it was implemented in the basal ganglia of non-human primates [13]. Our results motivate further research into optogenetic DBS technologies in humans as an effective alternative to electrical DBS.

Besides clinical applications, optogenetic stimulation is used as a research tool. Our results suggest that one of the reasons it may be more effective than electrical stimulation in control of beta-band synchronized oscillatory activity is that it does its job with less effective current injected into the neurons.

Finally, our results suggest that optogenetic stimulation may potentially be more efficacious than electrical stimulation at suppression of different types of oscillatory synchronized activity. Neural oscillations and synchrony are believed to be crucial for a variety of neural functions and dysfunctions [34]. Thus optogenetic stimulation may be more efficacious than electrical stimulation in controlling other types of neural synchrony, whether in some potential clinical applications or in experimental stimulation to control oscillations and synchrony-dependent functions.

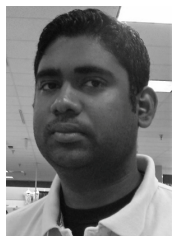
**Shivakeshavan Ratnadurai-Giridharan** received B.E. from Anna Univ., India (2007) and M.S. from Univ. of Florida, USA (2010) in electrical engineering. He received Ph.D. from Univ. of Florida, USA in Biomedical engineering (2010).

His specialization has been in computational neuroscience and neural engineering. He first received training in neuroscience and engineering at the Indian Institute of Technology, Madras (IITM), India (2007-2008), where he focused on modeling the Parkinsonian brain. He continued his training in these fields during his M.S. and Ph.D. studies with a focus on understanding the epilepsy. After graduation, he was a postdoctoral faculty in Indiana University Purdue University Indianapolis (2014-2016). There, his focus was on Parkinson's disease, beta synchrony, and feedback-based DBS for suppression of symptoms. Since 2016, he is a research associate at Burke Medical Research Institute (affiliated with Weill Cornell Medical College, at White Plains, NY). His focus is on building analytical tools and studying how electrical stimulation can alleviate motor symptoms after spinal cord injury.

Dr. Ratnadurai-Giridharan is a member of the Society for Neuroscience and has previously contributed to an IEEE article detailing how an electrooculogram system in conjunction with machine learning can help guide motorized wheel chairs for patients with paralysis.

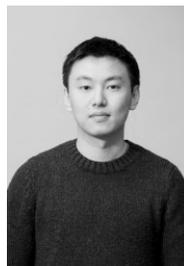
**Chung C. Cheung** received B.S. degree in mathematics from the University of Hong Kong, in 2008, and M.S. degree in mathematics from Purdue University, USA, in 2016. He is currently pursuing the Ph.D. degree in applied mathematics and statistics at Indiana University Purdue University Indianapolis, IN, USA. His research interests include computational neuroscience and big data analysis.

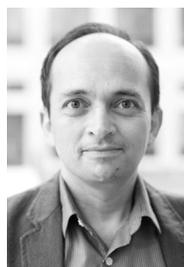
**Leonid L. Rubchinsky** received B.S. degree in physics from the University of Nizhny Novgorod, Russia, in 1994, M.S. degree in physics from the University of California, San Diego, in 1997, and Ph.D. in physics from the Institute of Applied Physics, Russian Academy of Science, in 2000.

He did a postdoctoral fellowship in computational neuroscience and clinical neurophysiology in the University of California, Davis, from 2001 to 2004. Since 2004 he is a faculty member (Associate Professor since 2010) at the Department of Mathematical Sciences, Indiana University Purdue University Indianapolis and at the Stark Neurosciences Research Institute, Indiana University School of Medicine. His research interests are in the areas of nonlinear dynamics, mathematical biology, and computational neuroscience, and include coupled oscillators and synchronization in living systems, dynamics and control of neuronal networks and assemblies, physiology of the basal ganglia in health and Parkinson's disease, brain rhythms and their synchronization, and neurophysiological data analysis.

Dr. Rubchinsky is presently serving on the Board of Directors of the Organization for Computational Neuroscience.